\documentclass[preprint, sort&compress, 5p]{elsarticle}
\usepackage{amsmath, amsthm, amssymb}
\usepackage[mathlines,displaymath]{lineno}
\usepackage[T1]{fontenc} 
\usepackage{graphicx} 
\usepackage[utf8]{inputenc} 
\usepackage[hyphens]{url}
\usepackage{bm}
\usepackage{enumerate} 
\usepackage[font=small,labelfont=bf]{caption}
\usepackage{chemformula}     
\usepackage[hidelinks,colorlinks,allcolors=blue]{hyperref}
\usepackage{cleveref}
\usepackage[color=lime]{todonotes}
\usepackage{placeins}
\usepackage{siunitx}
\usepackage{xspace}          
\usepackage{physics}

\presetkeys{todonotes}{inline}{}
\crefname{equation}{Eq.}{Eqs.}
\crefname{section}{Sec.}{Secs.}
\crefname{table}{Tab.}{Tabs.}
\crefname{figure}{Fig.}{Figs.}
\crefname{subfigure}{Fig.}{Figs.}

\usepackage{booktabs}

\newcommand*{\pd}[2]{\frac{\partial #1}{\partial #2}}

\newcommand*{\lp}{\left(}
\newcommand*{\rp}{\right)}

\newcommand*{\air}{{\textrm{air}}}

\newcommand*{\liq}{\ell}
\newcommand*{\vap}{{\textrm{g}}}

\newcommand*{\mbx}{\bm{x}}

\newcommand*{\mbu}{\bm{u}}
\newcommand*{\mbg}{\bm{g}}

\newcommand{\wat}{\ch{H2O}\xspace}
\newcommand{\watl}{\ensuremath{\ch{H2O}(\liq)}\xspace}
\newcommand{\watg}{\ensuremath{\ch{H2O}(\vap)}\xspace}
\newcommand{\nht}{\ch{NH3}\xspace}
\newcommand{\nhtl}{\ensuremath{\ch{NH3}(\textrm{aq})}\xspace}
\newcommand{\nhtg}{\ensuremath{\ch{NH3}(\vap)}\xspace}
\newcommand{\hh}{\ch{H2}\xspace}

\newcommand{\Sc}{\mathrm{Sc}}

\newcommand{\Prt}{\mathrm{Pr}}

\let\today\relax
\makeatletter
\def\ps@pprintTitle{%
    \let\@oddhead\@empty
    \let\@evenhead\@empty
    \def\@oddfoot{\footnotesize\itshape
         {Preprint -- 2024-04} \hfill\today}%
    \let\@evenfoot\@oddfoot
    }
\makeatother

\begin{document}

\begin{frontmatter}

\title{Influence of ammonia-water fog formation on ammonia dispersion from a liquid spill}

\address[SINTEF]{SINTEF Energy Research, Trondheim, NO-7465, Norway}
\author[SINTEF]{Hans Langva Skarsvåg\corref{c1}}
\cortext[c1]{Corresponding author}
\ead{hans.skarsvag@sintef.no}
\author[SINTEF]{Eirik Holm Fyhn}
\author[SINTEF]{Ailo Aasen}

\date{\today}

\begin{abstract}
  Ammonia is expected to play an important role in the green transition, both as a hydrogen carrier and a zero-emission fuel. The use of refrigerated ammonia is attractive due to its relatively high volumetric energy density and increased safety compared to pressurized solutions. Ammonia is highly toxic, and with new applications and increased global demand come stricter requirements for safe handling. Cold gaseous ammonia following a spill of refrigerated ammonia will in contact with humid air cause fog formation. In an environment rich in ammonia, these droplets will due to ammonia's strong hygroscopicity consist of considerable amounts of liquid ammonia as well as water. Fog formation affects the ammonia-air density and thus influences the dispersion dynamics, with a potentially significant impact on hazardous zones. In this work, we present a CFD model including an ammonia-water fog formation model based on accurate thermodynamics. This includes modeling the vapor-liquid equilibrium and accounting for the exothermic mixing of ammonia and water. We apply this CFD model to relevant cases and demonstrate the significant impact of the fog. We analyze the effect of varying relative humidity, fog visibility, influence of wind, and pool evaporation rate. Finally, we model the experimental Red Squirrel test 1F.
\end{abstract}

\begin{keyword}
Ammonia \sep Safety \sep Dispersion \sep CFD \sep Fog formation
\end{keyword}
\end{frontmatter}
\section{Introduction}
\label{sec:introduction}
A major challenge that must be overcome to facilitate a rapid green energy transition is to find suitable replacements for petroleum-based fuels.
Ammonia (\nht) is one candidate expected to play a major role.
Ammonia, like hydrogen (\hh), is a potential zero-emission fuel, either used directly for propulsion or as a hydrogen carrier~\citep{herbinet2022}.
In the former case, one must mitigate potential $\ch{NO_x}$ emissions~\citep{zhu2024}.
Like \hh~\citep{hydrogen_roadmap_iea,h2roadmap}, the demand for \nht is expected to increase dramatically in the future, with global demand of around 500~Mt by 2050~\citep{ammonia_roadmap_iea}.
Compared to \hh, \nht has around $\SI{70}{\percent}$ higher volumetric energy density~\citep{chatterjee2021}, and can be liquefied at much higher temperatures~\citep{cengel2011}.
Moreover, thanks to a century of ammonia used in agriculture, \nht transport is already a mature technology~\citep{royalsociety_ammonia}.

Ammonia is preferably stored in liquid form, due to the higher density.
This can be achieved either through high pressure, low temperature, or a combination of both.
Liquefaction from cooling at ambient pressure, so-called refrigerated \nht, is considered to be safer for storage than pressurized \nht~\citep{ng2023}.
This is achieved at \SI{-33.3}{\celsius}, which is significantly higher compared to the  \SI{-253}{\celsius}  required for \hh~\citep{cengel2011}.
Refrigerated \nht is safer not only because the high pressure results in a higher discharge rate, but also because a loss of containment (LOC) from a pressurized vessel will lead to rapid expansion, resulting in the formation of fine aerosols~\citep{nielsen1997,cleary2007,witlox2007,polanco2010}.
The resulting two-phase jet is heavy and can spread far close to the ground~\citep{kaiser1978,griffiths1982}.
On the other hand, the plume associated with the LOC of refrigerated \nht is buoyant and disperses more easily~\citep{dharmavaram2023}.
While \nht, unlike \hh, is not particularly flammable, it is acutely toxic.
Exposure even to relatively low concentrations may lead to severe irritation and eventually death \citep{atsdr2004}.
The worst example of this is the Dakar accident that happened in 1992, claiming 129 lives and injuring another 1150 people~\citep{dharmavaram2023_accident}.
It is therefore crucial to have good models to assess the risks associated with possible LOC under transportation or storage of \nht.

In the case of LOC of refrigerated \nht, cold liquid ammonia will spill, typically onto either ground or water.
In the latter case, about \SI{50}{\percent} of the spilled \nht will immediately evaporate, while the remaining \nht will dissolve into the water~\citep{dharmavaram2023}.
In the former case, cold \nht will spread on the ground and evaporate as heat is transferred from the ground.
As \nht vapor is lighter than air, it will generally rise.
However, as it rises and mixes with air it will also heat up, while the air is cooled.
The density of the cloud gradually reduces towards neutral buoyancy at complete dilution.
The mixture density is further affected by droplet formation. Cold air cannot hold the same absolute humidity as warm air, and 
cooling will eventually result in fog formation. Condensing droplets release latent heat, thus increasing the mixture temperature, and thus provides a positive contribution to the buoyancy. Moreover, \nht will mix into the water (\wat) droplets, giving rise to an ammonia-water fog. A model that aims at including these buoyancy effects must therefore be able to predict the amount of ammonia-water fog, and the corresponding energies released. This means also accounting for the exothermic reaction when water and ammonia mixes~\citep{renard2004}.
Additionally, the fog gives a visible indicator for the plume, although the toxic region can extend much further~\citep{dharmavaram2023} than the visible cloud.

The effect of fog formation from the dispersion of cold gas into a humid atmosphere has been studied for liquid \hh (L\hh) spills~\citep{liu2019,sun2024,giannissi2014,giannissi2019,shu2022}, liquefied natural gas (LNG)~\citep{giannissi2019,zhang2015,zhang2022,eberwein2020,sun2020,luo2018}, and \ch{CO2}~\citep{teng2021,liu2016}, where it has been shown that the effect can be significant, in particular for flows with relatively low velocities characterized by a small Froude number.
When refrigerated \nht is spilled onto either ground or water, the gaseous ammonia resulting from the evaporation will have a low velocity compared to the velocity of a jet coming from a pressurized tank.
Therefore, there are reasons to believe that the thermodynamic effects of fog formation can have an important impact on the dispersion of refrigerated ammonia.
Ammonia is very hygroscopic, which should enhance the thermodynamic fog effect due to interaction with water.

Experiments on pressurized ammonia release have been conducted in 1983~\citep{goldwire1985}, 1997~\citep{nielsen1997,bouet2004} and 2010~\citep{fox2011}.
More recently, experiments on the dispersion of refrigerated ammonia were conducted in 2022 at the Det Norske Veritas (DNV) Spadeadam site in Northern England~\citep{dharmavaram2023}.
There are also ongoing field tests with the release of ammonia, led by the U.S. Department of Homeland Security, under the project name Jack Rabbit III~\citep{JRIII-slides}.
From the experiments, it is clear that a visible fog can form even from the release of unpressurized ammonia~\citep{dharmavaram2023}.
The boiling rate is significantly higher for spills on water, but also in the case of slower evaporation from concrete, one can see the formation of fog near the source.

Two-phase ammonia dispersion has previously been modeled in a CFD framework for ammonia release from a pressurized vessel~\citep{labovsky2010}.
However, as the nature of liquid release from a pressurized vessel is quite different than for a refrigerated release, the model used in Ref.~\cite{labovsky2010} is not applicable in this case.
In particular, it does not include the liquid phase of water or the exothermic reaction between \wat and \nht.
On the other hand, the larger droplet sizes associated with a pressurized release motivate the use of a discrete particle modeling approach.

Here, we model the dispersion of ammonia coming from the LOC of refrigerated ammonia.
Both the vapor and droplets are included in an effective single-phase model, where the phase transitions are modeled by a relaxation model.
We also include the heat generated as \nht mixes with liquid \wat by effectively adding a nonlinear contribution to the latent heat.
This is more efficient compared to modelling the full set of chemical reactions between \nht and \wat.
We find that the formation of ammonia-water fog can have a significant effect on the dispersion of ammonia, reducing the distance it spreads by up to \SI{35}{\percent}.
It is therefore important to consider this effect, especially when evaluating experiments performed under humid conditions, as failure to do so can potentially lead to a model that underpredicts the potential extent of toxic ammonia under LOC.

We present the CFD model and the implementation of the ammonia-water fog model in \cref{sec:model}. In \cref{sec:results} we present results and discussion which both verify the numerics, the fog model and provide results on the fog visibility and the influence on the safety distance for various conditions. In addition to generic cases we apply our model to the Red Squirrel test 1F. Finally, we conclude our work in \cref{sec:conclusion}.

\begin{figure}[htpb]
  \centering
  \includegraphics[width=0.5\textwidth]{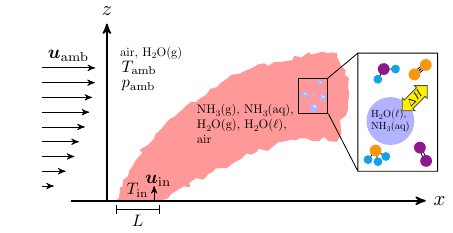}
  \caption{Illustration of the plume that develops in a crosswind following a spill of refrigerated ammonia.}
  \label{fig:sketch}
\end{figure}

\section{Model}\label{sec:model}

We consider the system sketched in \cref{fig:sketch}, where gaseous \nht at \SI{-33.15}{\degreeCelsius}, just above boiling temperature, is continuously released from a pool of area $L^2$.
As the cold ammonia spreads, it cools the ambient atmosphere, and water and ammonia may condense into droplets consisting of a binary mixture of the two substances.
To simplify calculations, we assume that these binary droplets follow the streamlines of the gas phase.
This assumption is reasonable in flows with low Stokes numbers, meaning that the particle relaxation time is small compared to the characteristic time of the flow.
The particle size in a fog is on the order of $\SI{1}{\micro\meter}$, and at most tens of $\si{\micro\meter}$~\citep{singh2011}.
This can be explained from a gap in the collision rate for drops of this size~\citep{poydenot2024}.
With a droplet size of $\SI{20}{\micro\meter}$ the relaxation time is around $\SI{1}{\milli\second}$, which we find is small compared to the characteristic time given by the velocity and curvature of the flow field.
A droplet of this size will also fall with a speed of about $\SI{10}{\milli\meter\per\second}$, which is small given the lifetimes of the droplets considered here.
Hence, the droplets can be assumed to follow streamlines.
With this, the model reduces to an effective single-phase mixture with five components: gaseous ammonia, \nhtg, liquid ammonia, \nhtl, water vapor, \watg, liquid water, \watl, and \air.
We write $\textrm{aq}$ to emphasize that this is not purely liquid \nht, since some of the \nht will react with water to form hydrated \ch{NH4+} and \ch{OH-} ions.
Naturally, this also means that \watl is not purely \wat molecules.

\subsection{Flow equations} \label{sec:flow_equations}
We model the dispersion using Reynolds-averaged transport equations~\citep{alfonso2009}.
In terms of turbulence-averaged variables, the transport equation for mass is
\begin{linenomath}\begin{equation}
  \pd{\lp\rho Y_i\rp}{t}  + \nabla\cdot \lp\rho\mbu Y_i\rp - \nabla\cdot \lp \frac{\mu_\text{eff}}{\Sc_t} \nabla Y_i\rp = R_i,
  \label{eq:Yeq}
\end{equation}\end{linenomath}
that of momentum transport is
\begin{linenomath}\begin{equation}
  \pd{\lp\rho \mbu\rp}{t}  + \nabla\cdot \lp\rho\mbu  \mbu\rp = \rho\mbg + \nabla\cdot \bar\sigma,
  \label{eq:Ueq}
\end{equation}\end{linenomath}
and the energy transport equation is
\begin{linenomath}\begin{equation}
  \pd{\lp\rho h\rp}{t}  + \nabla\cdot \lp\rho\mbu h\rp - \nabla\cdot \lp \alpha_\text{eff} \nabla h\rp = \pd{p}{t} + \dot{Q} + \mbu \cdot \nabla p.
  \label{eq:Eeq}
\end{equation}\end{linenomath}
Here, $\rho$ is density, $Y_i$ is the mass fraction, and the subscripts denote the component, meaning that
\begin{linenomath}\begin{equation}\label{eq:components}
i = \nhtg,\nhtl,\watg,\watl,\air.
\end{equation}\end{linenomath}
Moreover, $\mbu$ is velocity, $\mu_\text{eff}$ is the effective dynamic viscosity, including the contribution from turbulence, $\Sc_t$ is the turbulent Schmidt number, $\mbg$ is the acceleration due to gravity, $h$ is specific sensible enthalpy, meaning that heat of formation is not included, and $R_i$ is the rate associated with phase change.
That is, $R_{\nhtl} = -R_{\nhtg}$, $R_{\watl} = -R_{\watg}$ and $R_{\air} = 0$.
The implementation of the phase change and the associated release of latent heat through $\dot Q$ is detailed in \cref{sec:droplet}.

The stress tensor, $\bar\sigma$, includes the volumetric stress due to pressure, $p$, the viscous contribution to the deviatoric stress and the Reynolds stress, and is given by
\begin{linenomath}\begin{equation}
  \bar\sigma = -p\bar I + \mu_\text{eff} \left[\nabla\mbu + \lp\nabla\mbu\rp^T - \frac 2 3 \lp \nabla\cdot \mbu\rp\bar I\right] - \frac 2 3 k \bar I,
\end{equation}\end{linenomath}
where $\bar I$ is the identity tensor, $k$ is the turbulent kinetic energy, and where we have used the Boussinesq eddy viscosity assumption~\citep{alfonso2009}.
This means that the deviatoric contribution to the Reynolds stress is assumed proportional to the trace-less part of the Reynolds-averaged stress tensor, giving rise to $\mu_\text{eff} = \mu + \mu_t$, where $\mu$ is the dynamic viscosity and $\mu_t$ is the turbulent dynamic viscosity.
The thermal diffusion coefficient, $\alpha_\text{eff}$, is related to the dynamic viscosity through the Prandtl number, $\Prt$, and the turbulent Prandtl number, $\Pr_t$,
\begin{linenomath}\begin{equation}
  \alpha_\text{eff} = \frac{\mu}{\Prt} + \frac{\mu_t}{\Prt_t}.
\end{equation}\end{linenomath}
We apply the standard $k$-$\varepsilon$ model for turbulence~\citep{launder1974,sklavounos2004}, which means that
\begin{linenomath}\begin{equation}
  \mu_t = \rho C_\mu \frac{k^2}{\varepsilon},
\end{equation}\end{linenomath}
where $C_\mu$ is a constant and $\varepsilon$ is the turbulent kinetic energy dissipation rate.
The transport equations for $k$ and $\varepsilon$ read
\begin{linenomath}\begin{equation}
  \pd{\lp\rho k\rp}{t} + \nabla\cdot \lp \rho \mbu k \rp = \nabla\cdot\lp \rho D_k \nabla k \rp + P - \rho\varepsilon,
\end{equation}\end{linenomath}
and
\begin{multline}
  \pd{\lp\rho \varepsilon\rp}{t} + \nabla\cdot \lp \rho \mbu \varepsilon \rp = \nabla\cdot\lp \rho D_\varepsilon \nabla \varepsilon \rp 
  \\
  + \frac{C_1 \varepsilon}{k}\lp P + C_3 \frac 2 3 k \nabla\cdot \mbu \rp 
  - C_2 \rho\frac{\varepsilon^2}{k},
\end{multline}
respectively.
Here, $D_k$ and $D_\varepsilon$ are effective diffusion coefficients, $P$ is the turbulent kinetic energy production rate, and $C_1$, $C_2$, and $C_3$ are model coefficients.

We build on OpenFOAM v2206~\citep{weller1998}, and keep the default implementation of the $k$-$\varepsilon$ model where
\begin{subequations}
  \begin{align}
    D_k = &\frac{\mu_t}{\rho\sigma_k} + \frac{\mu}{\rho},
    \\ 
    D_\varepsilon = &\frac{\mu_t}{\rho\sigma_\varepsilon} + \frac{\mu}{\rho},
    \\
    P = &\mu_t \left\{\left[ \nabla  \mbu + \lp \nabla \mbu \rp^T  \right]\mathbin{:} \left[\nabla  \mbu\right] - \frac 2 3 \lp\nabla\cdot\mbu\rp^2\right\}
        -\frac 2 3 \rho k \nabla\cdot \mbu.
      \label{eq:turbkinenprod}
  \end{align}
\end{subequations}
We also use the default constant parameters, meaning that $\sigma_k = 1$, $\sigma_\varepsilon = 1.3$, $C_\mu = 0.09$, $C_1 = 1.44$, $C_2 = 1.92$, and $C_3 = 0$.

\subsection{Thermophysical properties}
The way OpenFOAM handles the thermophysical properties of mixtures is based on the assumption the properties of the mixture have the same functional form as those of the constituents and that the coefficients can be written in terms of the mass fractions and coefficients of the constituents.

As viscosity and thermal conductivity are expected to play a limited role, we set the viscosities and Prandtl numbers constant as functions of temperature, $T$, and pressure.
That is, the dynamic viscosity is given by
\begin{linenomath}\begin{equation}
  \mu = \sum_{i} Y_i \mu_i,
\end{equation}\end{linenomath}
where the sum runs over the components/phases in \Cref{eq:components} $\mu_{\nhtg}$, $\mu_{\watg}$, and $\mu_{\air}$ are taken from NIST Chemistry Webbook at $T=\SI{0}{\celsius}$~\citep{nist} and given in \cref{tab:properties}.
For simplicity, we also set $\mu_{\nhtl} = \mu_{\nhtg}$ and $\mu_{\watl} = \mu_{\watg}$.
Similarly, the Prandtl number is given by
\begin{linenomath}\begin{equation}
  \Prt = \sum_{i} Y_i \Prt_{i} ,
\end{equation}\end{linenomath}
where $\Pr_i$ are found in Ref.~\cite{nist} and given in \cref{tab:properties}.

Heat capacity is given by
\begin{linenomath}\begin{equation}
  c_p = \sum_{i} Y_i c_{pi},
  \label{eq:cp_mixing}
\end{equation}\end{linenomath}
when $c_p$ has dimensions $\si{\joule\per\kilogram\per\kelvin}$.
We set the heat capacities $c_{pi}$ fixed, as we are not able to improve the accuracy with temperature-dependent heat capacities while using the mixing rule specified by
\cref{eq:cp_mixing}.
The values for $c_{pi}$ at $T=\SI{0}{\degreeCelsius}$ we used are given in \cref{tab:properties}.
They were calculated with a highly accurate multiparameter equation of state (MEOS), which is the same one used in the REFPROP10 software~\citep{REFPROP10}, and was accessed through Thermopack~\citep{wilhelmsen2017}.
As specific heat is set constant, the specific sensible enthalpy, $h$ at temperature $T$ is
\begin{linenomath}\begin{equation}
  h = (T - T_\text{ref})c_{p},
\end{equation}\end{linenomath}
where $T_\text{ref} = \SI{293.15}{\kelvin}$ is a reference temperature.
How the total enthalpy is related to the sensible enthalpy is explained in \cref{sec:latent_heat}.

In terms of the pure component densities, $\rho_i(p, T)$, the inverse total density is given by
\begin{linenomath}\begin{equation}
  \rho^{-1} = \sum_{i} Y_i \rho^{-1}_i.
\end{equation}\end{linenomath}
We use the following density functions to describe liquids and the virial-corrected density of the gases
\begin{linenomath}\begin{equation}
  \rho_i^{-1}(p, T) =
  \begin{cases} 
    b_{0i},& \text{if } i=\text{liquid} \\
    b_{1i} T + b_{2i}\frac{T}{p}, & \text{if } i=\text{gas} 
  \end{cases}
  \label{eq:rPolynomial}
\end{equation}\end{linenomath}
where $b_{0i}$, $b_{1i}$, and $b_{2i}$ are constants.
For {\air} and \watg we use the ideal gas law, meaning that
\begin{linenomath}\begin{equation}
   b_{2i} = \frac{R}{W_i},
\end{equation}\end{linenomath}
where $R = \SI{8.314}{\joule\per\mol\per\kelvin}$ is the gas constant and $W_i$ is the molar mass of component $i$, while $b_{0i} = b_{1i} = 0$.
As there is some discrepancy from the ideal gas law for gaseous ammonia at low temperatures, we add the first virial correction to first order in temperature by letting $b_{0i}$ and $b_{1i}$ be nonzero.
In this case, we find $b_{0i}$ and $b_{1i}$ by fitting them to the virial expansion of the multiparameter equation of state~\citep{REFPROP10}.
The densities of \watl and \nhtl are set constant, meaning that only the values for $b_{0i}$ are nonzero.
The numerical values of $b_{0i}$, $b_{1i}$, and $b_{2i}$ are listed in \cref{tab:properties}.

\begin{table*}[tbp] 
\caption{Numerical values of constants} 
  \label{tab:properties} 
    \centering 
    \begin{tabular}{lSSSSS} 
    \toprule 
    {Constant} & {\nhtg} & {\watg} & \air & \nhtl & \watl\\ 
    \midrule 
    {$\mu_i$ ($10^{6}\,\si{\pascal\second}$)} & 9 & 9 & 17 & 9 & 9\\ 
    {$\Pr_i$ } & 0.82 & 1.00 & 0.71 & 1.80 & 2.28 \\ 
    {$c_{pi}$ ($10^{3}\,\si{\joule\per\kilogram\per\kelvin}$)} & 2.1 & 1.9 & 1 & 4.6 & 4.3\\ 
    {$b_{0i}$ ($10^{-3}\,\si{\meter\cubed\per\kilogram}$)} & -112 & 0 & 0 & 1.46 & 1.00 \\ 
    {$b_{1i}$ ($10^{-4}\,\si{\meter\cubed\per\kilogram\per\kelvin}$)} & 3.42 & 0 & 0 & 0 & 0\\ 
    {$b_{2i}$ (\si{\joule\per\kilogram\per\kelvin})} & 488.2 & 461.5 & 287.1 & 0 & 0\\ 
    \bottomrule 
\end{tabular} 
\\ 
\raggedright 
\end{table*}

\subsection{Atmospheric boundary layer}
Atmospheric turbulence plays a significant role in how the ammonia disperses into the atmosphere. 
From \cref{eq:turbkinenprod}, one can see that turbulent kinetic energy is produced by a term proportional to the deviatoric stress tensor, meaning that turbulence is generated from the shear stress.
It is therefore important to specify a non-uniform velocity profile for the domain boundaries.
Here we consider a logarithmic profile,
\begin{linenomath}\begin{equation}
  \mbu_\text{amb} = \frac{u^*}{\kappa} \ln \lp \frac {z + z_0}{z_0}\rp\mbx ,
  \label{eq:ABL_u}
\end{equation}\end{linenomath}
where $u^*$ is the friction velocity, $\kappa = 0.41$ is the von Kármán constant, $z_0$ is the aerodynamic roughness lenght, and $\mbx$ is the unit vector in the $x$-direction.
We choose $z_0 = \SI{2}{\centi\meter}$, corresponding roughly to heather moor or rough pasture~\citep{mcintosh1969}.
The friction velocity is chosen such that the wind velocity at $\SI{10}{\meter}$ is equal to a reference wind speed $u_\text{wind}$.

The logarithmic profile in \cref{eq:ABL_u} corresponds to a neutral boundary layer in equilibrium~\citep{yang2009,richards1993,hargreaves2007}, and the corresponding turbulent kinetic energy and turbulent kinetic dissipation rate are
\begin{linenomath}\begin{equation}
  k_\text{amb} = \frac{\lp v^*\rp^2}{\sqrt{C_\mu}}\sqrt{B_1 + B_2 u_\text{amb} \kappa / u^*},
  \label{eq:ABL_k}
\end{equation}\end{linenomath}
and
\begin{linenomath}\begin{equation}
  \varepsilon_\text{amb} = \frac{\lp v^*\rp^3}{(z + z_0)\kappa}\sqrt{B_1 + B_2 u_\text{amb} \kappa / u^*},
  \label{eq:ABL_eps}
\end{equation}\end{linenomath}
respectively.
The constants can be fitted to experimental data~\citep{yang2009}.
Using the default implementation, we choose $B_1 = 1$ and $B_2 = 0$.

That the atmosphere is neutrally stable means that the distribution of temperature as a function of height is adiabatic~\citep{obhukov1971}.
With only a small variation in the hydrostatic pressure over the domain, it is therefore a good approximation to set temperature constant for the incoming wind.
In atmospheric conditions that are not neutrally stable, turbulence is affected not only by surface roughness but also by buoyancy forces associated with the surface heat flux.
In stable or unstable conditions, one can use the Monin-Obukhov similarity theory~\citep{monin1954} to rewrite \cref{eq:ABL_u} to include a stability correction term~\citep{holtslag2014}.
When the suppression or enhancement of turbulence due to buoyancy is strong, it is also important to include buoyancy terms in the $k$-$\varepsilon$ model~\citep{vanmaele2006,kumar2014}.

\subsection{Latent heat} \label{sec:latent_heat}
The total absolute enthalpy density, $H$, is related to the specific sensible enthalpy, $h$, through
\begin{linenomath}\begin{equation}
  H = \rho h + \sum_{i } \rho_i h_{fi},
\end{equation}\end{linenomath}
where $h_{fi}$ is the specific heat of formation for component $i$.
Consider a phase change, where $\rho_i \mapsto \rho_i + \delta \rho_i$.
If the temperature is kept constant, the resulting change in enthalpy,
\begin{linenomath}\begin{equation}
  \delta H_l = \sum_{i} \delta \rho_i \left[\lp T-T_\text{ref}\rp c_{pi} + h_{fi}\right]
\end{equation}\end{linenomath}
is the latent heat.
The terms proportional to $c_{pi}$ are included in the sensible energy, while the terms proportional to the heat of formation are included in the source term $\dot Q$.
As $R_i$ is the rate of change in the mass of component $i$, the source term becomes
\begin{linenomath}\begin{equation}
  \dot Q = -\sum_{i } R_i h_{fi},
\end{equation}\end{linenomath}

As mentioned above, condensing a droplet of ammonia and water releases more heat than the heat that would be released if the binary droplet were made into two separate pure droplets.
This is because of the chemical reactions take place when $\nht$ mixes with $\wat$, and the hydrogen bonds between the molecules.
To incorporate this without keeping track of all the chemical reactions associated with the mixing, we let $h_{f\nhtl}$ and $h_{f\watl}$ depend on both $Y_{\nhtl}$ and $Y_{\watl}$.

\begin{figure}[htpb]
  \centering
  \includegraphics[width=0.48\textwidth]{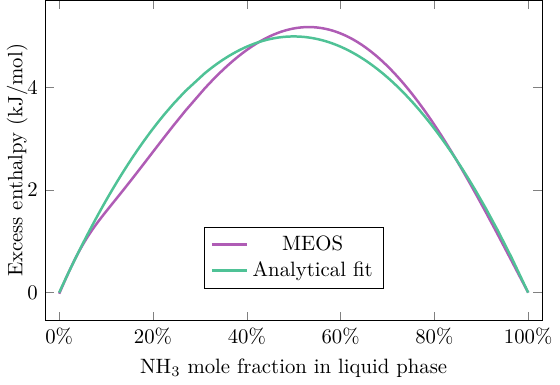}
  \caption{Excess enthalpy calculated assuming adiabatic mixing of air (\SI{20}{\degreeCelsius}) and \nht(g) (\SI{-33.15}{\degreeCelsius}). The result is not sensitive to the air temperature, and thus valid for all air temperatures used in this work.}
  \label{fig:excessEnthalpy}
\end{figure}

Let $\delta H_{\nht}$ and $\delta H_{\wat}$ be the latent heat associated with condensing \nhtl and \watl as pure droplets at $T = T_\text{ref}$.
The excess heat coming from the mixing of \nhtl and \watl is then the difference between the total latent heat, $\delta H_l$, and the latent heat that would be released from the same amount of pure \nht and \wat droplets.
That is,
\begin{linenomath}\begin{equation}
  \delta H_\text{excess} = \delta H_l - \delta H_{\nht} - \delta H_{\wat}.
\end{equation}\end{linenomath}
We use a simple analytical fit for the excess enthalpy,
\begin{linenomath}\begin{equation}
  \delta H_\text{excess} = -4 \delta H_0^\text{excess} \frac{n_{\nhtl} n_{\watl}}{n_{\nhtl} + n_{\watl}},
  \label{eq:H_excess}
\end{equation}\end{linenomath}
where $n_{\nhtl}$ and $n_{\watl}$ are the number of $\nht$ and $\wat$ molecules per volume in the liquid phase before mixing, respectively, and $\delta H_0^\text{excess}$ is a constant.
This gives a straighforward implementation in the CFD model and, as seen in \cref{fig:excessEnthalpy}, $\delta H_0^\text{excess} = \SI{5}{\kilo\joule\per\mole}$ gives a resonable fit to the values obtained from the accurate multiparameter equation of state~\citep{REFPROP10}.

To incorporate the excess enthalpy into $\dot Q$, one must differentiate \cref{eq:H_excess} with respect to time.
We find that
\begin{multline}
  \pd{\lp\delta H_\text{excess}\rp}{t} = -\frac{4 \delta H_0^\text{excess} n_{\watl}^2}{(n_{\nhtl} + n_{\watl})^2W_{\nht}} R_{\nhtl}
  \\
  -\frac{4 \delta H_0^\text{excess} n_{\nhtl}^2}{(n_{\nhtl} + n_{\watl})^2W_{\wat}}R_{\watl},
\end{multline}
where $W_{\nht}$ and $W_{\wat}$ are the molar masses of $\nht$ and $\wat$, respectively.
To incorporate the latent heat, including the excess enthalpy, we therefore set 
\begin{linenomath}\begin{equation}
  h_{f\watl} = -\SI{2443}{\kilo\joule\per\kilogram} - \frac{4 \delta H_0^\text{excess} n_{\nhtl}^2}{(n_{\nhtl} + n_{\watl})^2W_{\wat}}
\end{equation}\end{linenomath} 
and 
\begin{linenomath}\begin{equation}
  h_{f\nhtl} = -\SI{1229}{\kilo\joule\per\kilogram} - \frac{4 \delta H_0^\text{excess} n_{\watl}^2}{(n_{\nhtl} + n_{\watl})^2W_{\nht}},
\end{equation}\end{linenomath}
where the numerical values at $T_\text{ref}$ in the absence of mixing enthalpy are extrapolated from Ref.~\cite{nist}.

\subsection{Droplet model} \label{sec:droplet}
The phase-change rates, $R_{\nhtl}$ and $R_{\watl}$, are chosen to facilitate rapid equilibration. The assumption here being
that the gas-droplet mixture is close to local thermodynamic equilibrium.
To this end we compute the equilibrium concentrations, $n_{\nhtl}^\textrm{eq}$ and $n_{\watl}^\textrm{eq}$, and define
\begin{linenomath}\begin{equation}
  R_{\nhtl} =
    \frac{W_{\nht}k_0 (n_{\nhtl}^\textrm{eq} - n_{\nhtl})n_{\nhtl}}{n_{\nhtl} + n_{\nhtg}}
\end{equation}\end{linenomath}
when $n_{\nhtl} > n_{\nhtl}^\textrm{eq}$, and
\begin{linenomath}\begin{equation}
  R_{\nhtl} =
    \frac{W_{\nht}k_0 (n_{\nhtl}^\textrm{eq} - n_{\nhtl})n_{\nhtg}}{n_{\nhtl} + n_{\nhtg}}
\end{equation}\end{linenomath}
when $n_{\nhtl} < n_{\nhtl}^\textrm{eq}$.
Here, $k_0$ is a constant, and $n_{\nhtg}$ is the number of $\nht$ molecules per volume in the gas phase.
Similarly, $n_{\watg}$ is the number of $\wat$ molecules per volume in the gas phase.
The  phase-change rate associated with the \wat-phase transition is defined as
\begin{linenomath}\begin{equation}
  R_{\watl} =
    \frac{W_{\wat}k_0 (n_{\watl}^\textrm{eq} - n_{\watl})n_{\watl}}{n_{\watl} + n_{\watg}}
\end{equation}\end{linenomath}
when $n_{\watl} > n_{\watl}^\textrm{eq}$, and
\begin{linenomath}\begin{equation}
  R_{\watl} =
    \frac{W_{\wat}k_0 (n_{\watl}^\textrm{eq} - n_{\watl})n_{\watg}}{n_{\watl} + n_{\watg}}
\end{equation}\end{linenomath}
when $n_{\watl} < n_{\watl}^\textrm{eq}$.

The constant $k_0$ is chosen large such that the system is locally close to vapor-liquid equilibrium.
We use OpenFOAM's framework for chemical reactions for the phase change. The phase change then uses an ordinary differential equation solver to solve the phase-change dynamics in each time step, meaning that one can choose a large value for $k_0$ without having to solve the flow dynamics with correspondingly small time steps.
Here we choose $k_0 = \SI{100}{\per\second}$.

\begin{figure}[htpb]
  \centering
  \includegraphics[width=0.48\textwidth]{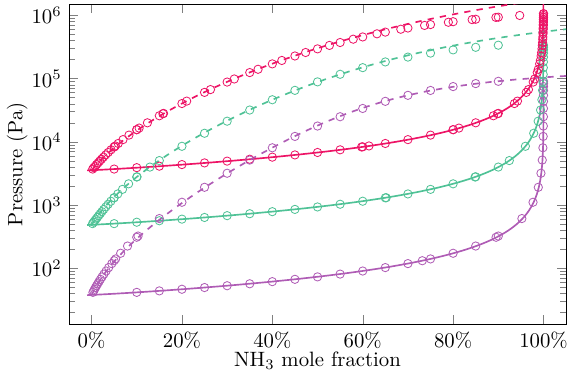}
  \caption{Correlation for the phase envelope (curves) and pseudoexperimental data (markers).
    Dashed lines show \nht mole fractions in the liquid phase, $x_{\nht}^\textrm{eq}$, and solid lines show \nht mole fractions in the vapor phase, $y_{\nht}^\textrm{eq}$.
  The colors correspond to different temperatures, where red is \SI{240}{\kelvin}, green is \SI{270}{\kelvin}, and purple is \SI{300}{\kelvin}.}
  \label{fig:VLEcorrelation}
\end{figure}

To determine $n_{\nhtl}^\textrm{eq}$ and $n_{\watl}^\textrm{eq}$ we use a correlation function for the phase envelope.
This makes the phase equilibrium computations considerably faster than having to do a full flash calculation to find the two-phase equilibrium.
The correlation function for the liquid equilibrium composition is
\begin{linenomath}\begin{multline}
   x_{\nht}^\textrm{eq} = \frac{1}{1+ a_1 \ln\lp p_r\rp}\bigg[\lp a_2 + a_3 T + a_4 T^2\rp \ln\lp p_r \rp 
   \\ + a_5 \ln\lp p_r\rp^2 + a_6 \ln\lp p_r\rp^3 + a_7 \ln\lp p_r\rp^4 \biggr],
\end{multline}\end{linenomath}
where $p_r$ is the partial pressure of the $\nht$ and $\wat$ subsystem normalized by the $\wat$ saturation pressure, and
\begin{linenomath}\begin{equation}
  x_{\nht}^\textrm{eq} = \frac{n_{\nhtl}^\textrm{eq}}{n_{\nhtl}^\textrm{eq} + n_{\watl}^\textrm{eq}}.
\end{equation}\end{linenomath}
The corresponding correlation function for the vapor equilibrium composition is
\begin{linenomath}\begin{multline}
   y_{\nht}^\textrm{eq} = \tanh\bigg[\lp d_1 + d_2 T\rp \ln\lp p_r \rp 
   \\ + d_3 \ln\lp p_r\rp^{g_1} + d_4 \ln\lp p_r\rp^{g_2} \biggr],
\end{multline}\end{linenomath}
where
\begin{linenomath}\begin{equation}
  y_{\nht}^\textrm{eq} = \frac{n_{\nhtg}^\textrm{eq}}{n_{\nhtg}^\textrm{eq} + n_{\watg}^\textrm{eq}}.
\end{equation}\end{linenomath}
The coefficients are shown in \cref{tab:correlation}.
The maximum error in the ammonia fraction is 0.0039 in the liquid phase and 0.011 in the vapor phase.
The accuracy is illustrated in Fig. \ref{fig:VLEcorrelation}.

Defining the total \nht molar composition fraction,
\begin{linenomath}\begin{equation}
  z_{\nht} = \frac{n_{\nhtg} + n_{\nhtl}}{n_{\nhtg} + n_{\nhtl} + n_{\watg} + n_{\watg}},
\end{equation}\end{linenomath}
and using the fact that $z_{\nht} = z_{\nht}^{\textrm{eq}}$, one can solve $z_{\nht}^{\textrm{eq}} = \beta^\ell x_{\nht}^{\textrm{eq}} + (1-\beta^\ell)y_{\nht}^{\textrm{eq}}$ for the liquid mole fraction $\beta^\ell$,
\begin{linenomath}\begin{equation}
  \beta^\ell = \frac{z_{\nht} - y_{\nht}^{\textrm{eq}}}{x_{\nht}^{\textrm{eq}} - y_{\nht}^{\textrm{eq}}}.
  \label{eq:beta}
\end{equation}\end{linenomath}
This, in turn, gives the equilibrium concentrations, $n_{\nhtl}^{\textrm{eq}} = x_{\nht}^\textrm{eq}\beta^\ell\sum_{i}n_i$ and $n_{\watl}^{\textrm{eq}} = (1 - x_{\nht}^\textrm{eq})\beta^\ell\sum_{i}n_i$.

\begin{table}[tbp] 
\caption{Correlation coefficients.} 
  \label{tab:correlation} 
    \centering 
    \begin{tabular}{lS} 
    \toprule 
    {Coefficient} & {Value} \\ 
    \midrule 
    {$a_1$} & -1.172e-1 \\ 
    {$a_2$} & 8.569e-2 \\ 
    {$a_3$ (\si{\per\kelvin})} & -7.159e-4 \\ 
    {$a_4$ (\si{\per\kelvin\squared})} & 1.986e-6 \\
    {$a_5$} & 7.807e-3 \\ 
    {$a_6$} & -1.861e-3 \\ 
    {$a_7$} & 7.144e-5 \\
    {$d_1$} & -1.461e2 \\ 
    {$d_2$ (\si{\per\kelvin})} & 2.992e-4 \\ 
    {$d_3$} & 1.468e2 \\ 
    {$d_4$} & 6.788e-3 \\ 
    {$g_1$} & 9.991e-1 \\ 
    {$g_2$} & 2.922 \\
    \bottomrule 
\end{tabular} 
\end{table} 

\Cref{eq:beta} becomes numerically inaccurate in regions with negligible amounts of $\nht$, as both the numerator and the denominator tend to $0$.
To get the correct behavior in the limit $z_{\nht}\to 0$, we shift to a pure water droplet model where $R_{\nhtl} = 0$ and the amount of water droplet is determined by having the partial pressure of $\watg$ bounded by the saturation pressure, $p_{\wat}^\text{sat.}(T)$.
That is, $n_{\watg}^{\text{eq}} = \min(p_{\wat}^\text{sat.}/p, n_{\wat})$ and $n_{\watl}^{\text{eq}} = n_{\wat} - n_{\watg}^{\text{eq}}$, where $n_{\wat} = n_{\watg} + n_{\watl}$.
We use a linear weight function to smoothly transition between the pure water droplet model when $x_{\nht}^\text{eq} < \SI{0.3}{\percent}$ and the full binary droplet model at $x_{\nht}^\text{eq} > \SI{2}{\percent}$.

\subsection{Mesh}
\begin{figure}[htpb]
  \centering
  \includegraphics[width=0.5\textwidth]{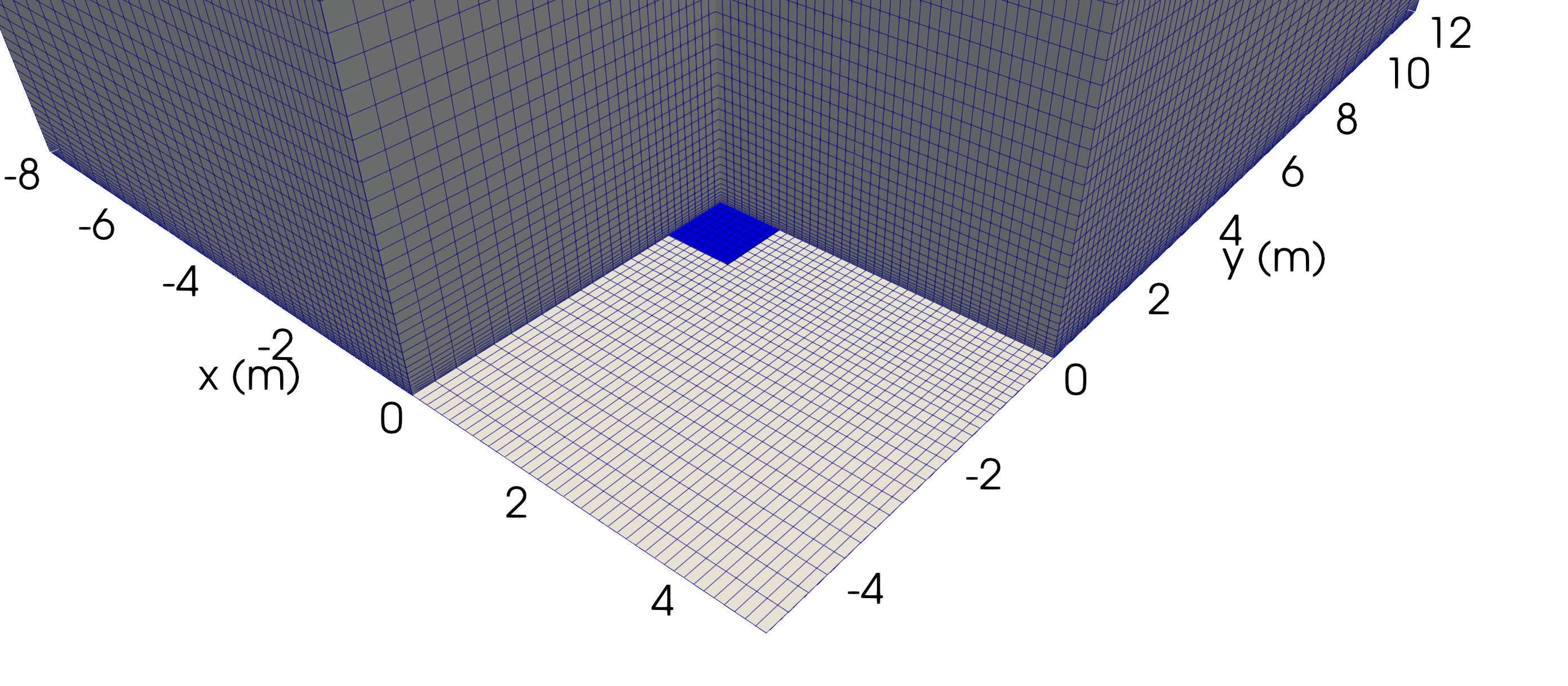}
  \caption{Near-source region of mesh used in all simulations. The blue square is one quadrant of the \SI{2}{\meter}$\times$\SI{2}{\meter} source inlet used in most simulations.}
  \label{fig:mesh}
\end{figure}

To model the system sketched in \cref{fig:sketch}, we use a mesh with dimensions $\SI{182}{\meter} \times \SI{24}{\meter} \times \SI{24}{\meter}$.
As shown in \cref{fig:mesh}, we grade the mesh such that it has a higher resolution where there are steeper gradients near the plume origin at $(0, 0, 0)$.
We are mostly interested in the ammonia concentrations below a few meters.
As a result, we have partitioned the domain into two regions, where the coarse upper region serves as a buffer to the far field region at $z > \SI{24}{\meter}$ where we fix the velocity, the turbulent kinetic energy, and the turbulent kinetic dissipation rate according to \cref{eq:ABL_u,eq:ABL_k,eq:ABL_eps}.
We also fix these values at $y = \pm\SI{12}{\meter}$.
This is done to ensure that the solution goes asymptotically to the atmosphere boundary layer solution also for large $x$.
The boundary condition is justified for the release rates we are considering here, as the plume is sufficiently diluted at these distances from the source.
The grid is also graded near the ground to ensure an appropriate value of the dimenensionless wall distance, $y^+$, between $30$ and a few hundred.

\section{Results and discussion}\label{sec:results}%
Having introduced the model in the previous section, we next turn to application and numerical verification.
Before investigating how the formation of ammonia-water fog affects the dispersion of ammonia, we first numerically test the underlying assumptions.
This is done in two parts.
First, we perform an error estimation based on a convergence test to determine a reasonable grid resolution.
Then, we test the droplet model by comparing results to analytical calculations.

To simplify the presentation of the parameter study, we define a \emph{reference case} in \cref{tab:ref_case}.
For most of the results presented in the following, we take the reference case as a starting point and vary one or more of the parameters.
The ambient temperature of \SI{30}{\celsius} was chosen because we anticipated that buoyancy from fog would be more relevant for higher temperatures.
However, we surprisingly found a strong effect also at lower temperature, as we show in \cref{sec:wind}.
A wind speed of \SI{3}{\meter\per\second} was chosen as this is a typical wind speed close to where one might expect the ammonia to spread the farthest, and ammonia inlet size and speed was chosen to be characteristic of the mass rate in the initial phase of \nht spill onto land~\citep{dharmavaram2023}.

\begin{table}[tbp] 
  \caption{Definition of the \emph{reference case}, where $T_\text{amb}$ is the ambient temperature, $u_\text{wind}$ is the wind speed measured at an altitude of \SI{10}{\meter}, $u_\text{in}$ is the ammonia inlet speed, and $L$ is the ammonia inlet size.} 
  \label{tab:ref_case} 
    \centering 
    \begin{tabular}{cccc} 
    \toprule 
    {$T_\text{amb}$} & {$u_\text{wind}$} & {$u_\text{in}$} & $L$\\ 
    \midrule 
    {\SI{30}{\celsius}} & {\SI{3}{\meter\per\second}} & {\SI{0.1}{\meter\per\second}} & \SI{2}{\meter} \\ 
    \bottomrule 
\end{tabular} 
\\ 
\raggedright 
\end{table}

\subsection{Grid refinement}
To determine the convergence of the solution, we studied the variation of safety distance as a function of the number of cells.
Going from 1.1 million cells to 7.3 million cells changed the safety distances by less than \SI{3}{\percent} at a wind speed of \SI{3}{\meter\per\second}.
Due to limitations on computational resources we were unable to go beyond this and establish a clear indication of convergence.
To compensate, we do a grid-convergence study on a two-dimensional grid, where we use the discretization of the $y=0$ plane of the 3D mesh.

\begin{figure}[htpb]
  \centering
  \includegraphics[width=0.48\textwidth]{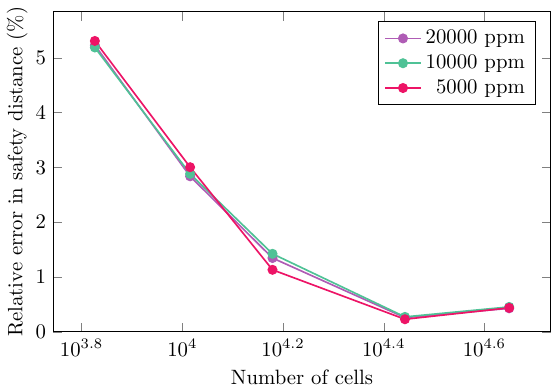}
  \caption{Relative error in safety distances as a function of the number of cells in a two-dimensional grid. The simulations are run with an ammonia inlet size of $L = \SI{0.5}{\meter}$, an ammonia inlet velocity of $u_\text{in} = \SI{0.05}{\meter\per\second}$, a relative humidity of \SI{99.5}{\percent} and remaining parameters given by \cref{tab:ref_case}. The errors are estimated by taking the difference with the results obtained from a finer mesh with $10^{4.8}$ cells.}
  \label{fig:convergence2D}
\end{figure}

The two-dimensional convergence study is shown in \cref{fig:convergence2D}.
In the absence of an analytical solution, the errors are estimated from the deviations from the safety distance computed with $10^{4.8}$ cells.
This would correspond to 23 million cells in the three-dimensional model.
From \cref{fig:convergence2D}, we see that a good compromise between accuracy and convergence can be found around $10^{4.2}$ cells, where the error in safety distance is around \SI{1}{\percent}.
Converting this mesh to the three-dimensional case yields 1.1 million cells.

We expect a slower convergence in three dimensions because the plume has a more complicated shape, so the error at 1.1 million cells is likely on the order of \SI{3}{\percent}.
Nevertheless, as we see a similar rate of convergence in 3D both with and without humidity, we expect the discretization error to be approximately independent of humidity.
We use the mesh with 1.1 million cells in all simulations below.

\subsection{Adiabatic mixing}\label{subsec:admix}
Adiabatic mixing would occur at constant pressure in the absence of heat diffusion.
In this case the temperature of a parcel of fluid would be determined solely by the constituents.
While this is not true here, it is a good approximation.
Therefore, we can use this observation to test the implementation of the droplet model.
That is, we can compute, for example, density as a function of ammonia mass fraction under the assumption of adiabatic mixing, and compare this with the values obtained at the same ammonia fraction in the CFD implementation.
One should not expect an exact match at every point, but it gives a good indication of whether the model is implemented correctly and whether the system is able to reach local equilibrium.

\begin{figure}[htpb]
  \centering
  \includegraphics[width=0.5\textwidth]{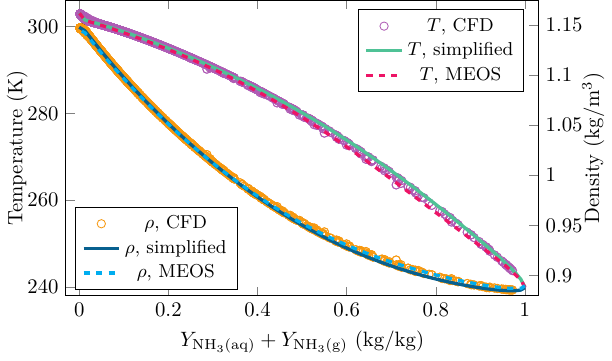}
  \caption{Temperature, $T$, and density, $\rho$, calculated using the full multiparameter EoS (dashed), simplified thermodynamics (solid) and the simplified thermodynamics implemented in the CFD model (circles).}
  \label{fig:admix_T_rho}
\end{figure}

\Cref{fig:admix_T_rho} shows the temperature and density computed as a function of $Y_{\nhtl} + Y_{\nhtg}$ under the assumption of adiabatic mixing as well as the corresponding values extracted from the CFD model for the reference case (\cref{tab:ref_case}) with a relative humidity of $\SI{90}{\percent}$.
The CFD model was run to steady state and all the cell values along the slice $y = 0$ were extracted, while the theoretical values were calculated using both the full multiparameter EoS and the simplified approximation that is implemented in the CFD model.
There is excellent agreement between all three sets of densities and temperatures.
This indicates both that the droplet model is able to reach an approximate equilibrium and that the simplified model used here is a good approximation to the full EoS.

\begin{figure}[htpb]
  \centering
  \includegraphics[width=0.45\textwidth]{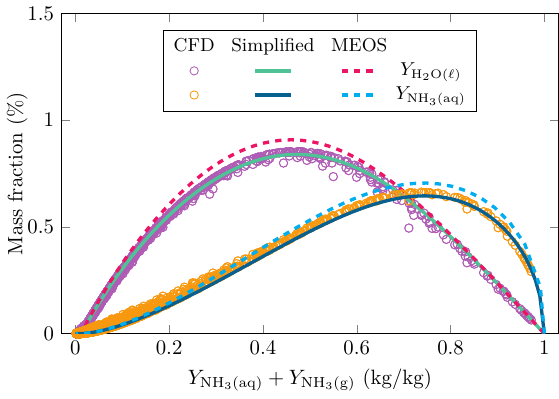}
  \caption{Droplet mass fractions, $Y_{\watl}$ and $Y_{\nhtl}$ calculated using the full multiparameter EoS (dashed), simplified thermodynamics (solid) and the simplified thermodynamics implemented in the CFD model (circles).}
  \label{fig:admix_droplets}
\end{figure}

\Cref{fig:admix_droplets} shows the mass fractions of the droplet constituents, $Y_{\watl}$ and $Y_{\nhtl}$ as a function of $Y_{\nhtl} + Y_{\nhtg}$ and under the same conditions as \cref{fig:admix_T_rho}.
\Cref{fig:admix_droplets} also shows a good match overall, especially between the CFD results and the adiabatic results using the simplified model.
Note that in the end it is the effect on density that influences the CFD results for dispersion.

\subsection{Fog visibility}
\begin{figure}[htpb]
  \centering
  \includegraphics[width=0.48\textwidth]{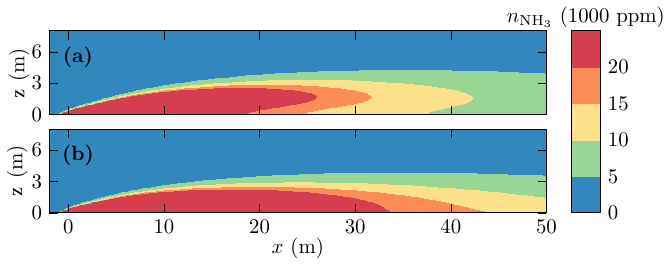}
  \caption{Steady-state \nht concentration, $n_{\nht}$ for the reference case (\cref{tab:ref_case}) with a relative humidity of (a) $\SI{99.5}{\percent}$ and (b) $\SI{50}{\percent}$.}
  \label{fig:CFD-illustration-ppm}
\end{figure}

\begin{figure}[htpb]
  \centering
  \includegraphics[width=0.48\textwidth]{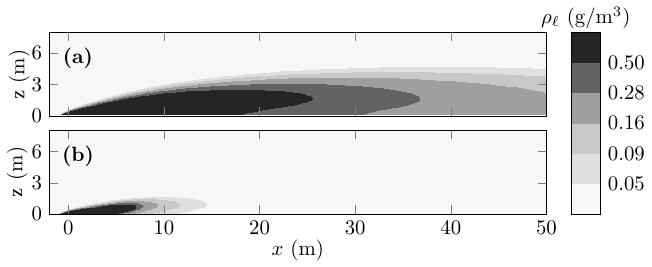}
  \caption{Steady-state fog density, $\rho_\ell = \rho_{\nhtl} + \rho_{\watl}$, for the reference case (\cref{tab:ref_case}) with a relative humidity of (a) $\SI{99.5}{\percent}$ and (b) $\SI{50}{\percent}$.}
  \label{fig:CFD-illustration-fog}
\end{figure}

The fog that forms during a release is the only visual indication of where the dangerous cloud is moving. This can aid escaping personnel in avoiding exposure. At the same time,
cases with little fog development can lead to a false sense of security. In \cref{fig:CFD-illustration-ppm,fig:CFD-illustration-fog} we show the steady-state \nht and fog concentration for the reference case (\cref{tab:ref_case}) with \SI{50}{\percent} and \SI{99.5}{\percent} relative humidity (RH). As expected, more fog is formed in the high-humidity case. Typically, a fog density of \SI{0.5}{g\per \meter\cubed} (droplet mass per volume air) is considered dense, with a visibility of about \SI{50}{\meter} and \SI{0.05}{\gram \per \meter\cubed} is considered a medium fog, with a visibility of about \SI{300}{\meter}~\citep{mao2004}. Note that liquid \nht has approximately the same refractive index as water, so we expect that liquid \nht and water have about the same contribution to visibility. For the high-humidity case, there is a reasonable overlap between the visible cloud and the regions of dangerous concentration. This is drastically different from the \SI{50}{\percent} RH case, where the visible cloud extends only to \nht concentrations of about 100\,000~ppm.

\begin{figure}[htpb]
  \centering
  \includegraphics[width=0.48\textwidth]{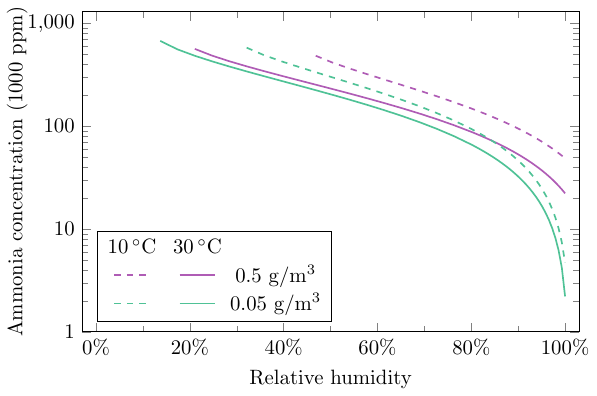}
  \caption{\nht concentration as a function of ambient humidity for dense (\SI{0.5}{\gram\per\kg}) and thin (\SI{0.05}{\gram\per\kg}) fog threshold at ambient temperatures of \SI{10}{\celsius} (dashed) and \SI{30}{\celsius} (solid) under the assumption of adiabatic mixing.}
  \label{fig:C_vs_RH-fog}
\end{figure}

As shown in Subsection \ref{subsec:admix} the mixing is approximately adiabatic. By using this approximation we can quantify the relation between fog visibility and \nht concentration. Given an ambient temperature and RH, we can then calculate the droplet concentration at a given \nht concentration. The result is shown in \cref{fig:C_vs_RH-fog}, where we plot the \nht concentration that yields dense fog (\SI{0.5}{\gram\per\kg}) and thin fog (\SI{0.05}{\gram\per\kg}), for ambient temperature $T=\SI{10}{\celsius}$ and $T=\SI{30}{\celsius}$. There is a strong dependence on relative humidity, and only for the most humid environments does the fog extend as far as the lower dangerous concentrations. This means that one should not under any circumstance equate the absence of fog to safe concentration levels. 

\subsection{Parameter study}
To characterize the spread of ammonia, we define a \emph{safety distance} which is a function of ammonia concentration.
The safety distance, $x_\text{sd}$, at a given ammonia concentration is the minimum distance at which the maximum concentration is below that threshold concentration for all positions below $z=\SI{3}{\meter}$ altitude. We believe this choice is a bit more versatile than simply choosing the concentration at a certain height, but in the end we do not expect the results to be very sensitive to this definition.


\subsubsection{Humidity}

\begin{figure}[htpb]
  \centering
  \includegraphics[width=0.48\textwidth]{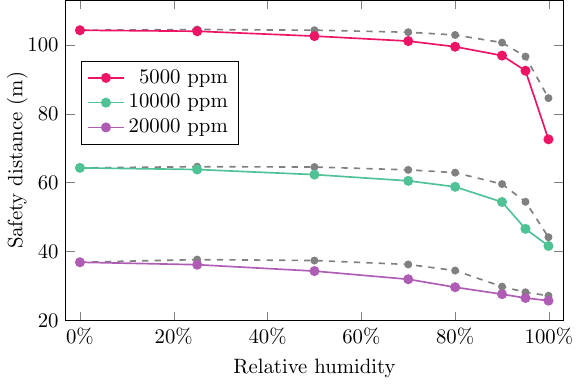}
  \caption{Safety distances as a function of relative humidity in the reference case given by \cref{tab:ref_case}. Gray dashed lines show the results with only a pure water droplet model.}
  \label{fig:humidity}
\end{figure}

\begin{figure}[htpb]
  \centering
  \includegraphics[width=0.48\textwidth]{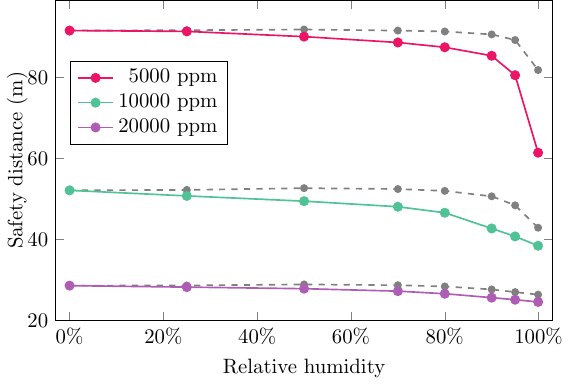}
  \caption{Safety distances as a function of relative humidity with ambient temperature $T_\text{amb} = \SI{10}{\celsius}$. The rest of the parameters are as in the reference case given by \cref{tab:ref_case}. Gray dashed lines show the results with only a pure water droplet model.}
  \label{fig:humidity_T10}
\end{figure}

As seen in \Cref{fig:CFD-illustration-ppm,fig:CFD-illustration-fog}, fog-induced buoyancy lifts the plume in a humid environment, giving a much shorter spread of \nht. \Cref{fig:humidity,fig:humidity_T10} show the safety distance for three different ammonia concentrations as a function of relative air humidity with a wind speed of \SI{3}{\meter\per\second} and with ambient temperatures of \SI{30}{\celsius} and \SI{10}{\celsius}, respectively.
As seen in \Cref{fig:CFD-illustration-ppm,fig:CFD-illustration-fog}, fog-induced buoyancy lifts the plume in a humid environment, giving a much shorter spread of \nht. \Cref{fig:humidity} shows the safety distance for three different ammonia concentrations as a function of relative air humidity at \SI{30}{\celsius} and a wind speed of \SI{3}{\meter\per\second} at \SI{10}{\meter} altitude.
At 20000~ppm, 10000~ppm and 5000~ppm, the safety distance is reduced by \SI{30}{\percent}, \SI{35}{\percent}, and \SI{30}{\percent}, respectively, when going from \SI{0}{\percent} to \SI{99.9}{\percent} relative humidity.
That the safety distance is reduced is reasonable, since increasing air humidity leads to more fog formation which enhances the buoyancy of the plume.

The gray dashed lines in \cref{fig:humidity,fig:humidity_T10} show the results with only \wat droplets, where the concentration of droplet is found from the saturation pressure, as explained in \cref{sec:droplet}.
The simpler pure \wat droplet model has a qualitatively similar response to ambient humidity as the full binary droplet model, but the effect is considerably smaller.

Relative humidity of \SI{0}{\percent} is not physical, but is included to show the result that would be obtained from not including the effect of fog formation in the CFD model.
From \cref{fig:humidity} one can see that, especially for the lower concentrations, the effect of fog only becomes strong near around \SI{80}{\percent} relative humidity.
Therefore, a fogless CFD simulation would likely suffice to model ammonia dispersion on a relatively dry day with for example \SI{60}{\percent} relative humidity.
On the other hand, simulating the dispersion of ammonia on a humid day can overpredict the safety distance by as much as \SI{54}{\percent} at 10000~ppm.
According to the ATSDR~\citep{atsdr2004}, ammonia exposure is rapidly fatal at \numrange{5000}{10000}~ppm.
Therefore, if empirical models~\citep{bricard1998,banerjee1996} are created based on experiments taken on a humid day, not considering how the ammonia-water for affected the experiments could lead to underpredicting the regions with potentially fatal amounts of \nht by more than \SI{50}{\percent}.

\begin{figure}[htpb]
  \centering
  \includegraphics[width=0.48\textwidth]{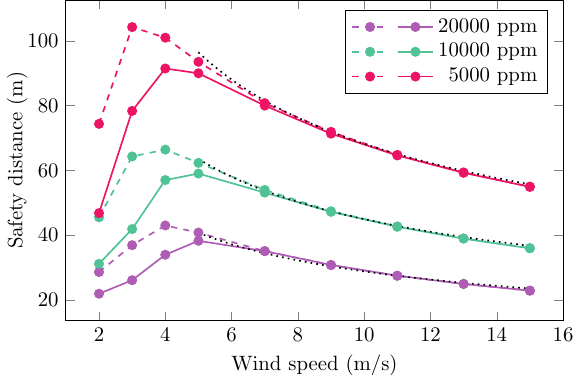}
  \caption{Safety distances as a function of wind speed, $u_\text{wind}$, in the reference case (\cref{tab:ref_case}) with relative humidity equal to \SI{99.5}{\percent} (solid) and \SI{0}{\percent} (dashed). The black dotted lines are proportional to $1/\sqrt{u_\text{wind}}$}
  \label{fig:T_30_vs_wind}
\end{figure}

\begin{figure}[htpb]
  \centering
  \includegraphics[width=0.48\textwidth]{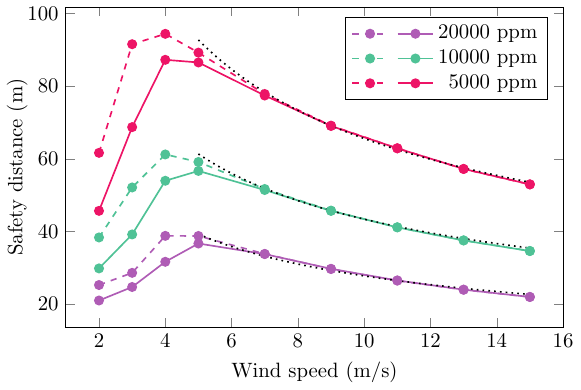}
  \caption{Safety distances as a function of wind speed, $u_\text{wind}$, in the reference case (\cref{tab:ref_case}) with ambient temperature $T_\text{amb} = \SI{10}{\celsius}$ and relative humidity equal to \SI{99.5}{\percent} (solid) and \SI{0}{\percent} (dashed). The black dotted lines are proportional to $1/\sqrt{u_\text{wind}}$}
  \label{fig:T_10_vs_wind}
\end{figure}

\subsubsection{Wind speed}\label{sec:wind}
\Cref{fig:T_30_vs_wind,fig:T_10_vs_wind} show safety distance as a function of wind speed with ambient temperature equal to \SI{30}{\celsius} and \SI{10}{\celsius}, respectively.
Again, we see that the safety distance is reduced in the presence of fog formation.
Importantly for safety analysis, it is not only the safety distance at a given wind speed that is reduced, but also the maximal safety distance over all wind speeds is also reduced by up to \SI{25}{\percent}.
For real applications, one often only has statistical knowledge of the wind data, and the wind is rarely constant in space or time.
Nevertheless, the results presented here indicate that the worst-case scenario on a humid day is less severe than the worst-case scenario on a dry day by up to as much as \SI{25}{\percent}.

From \cref{fig:T_30_vs_wind,fig:T_10_vs_wind} one can observe two different regimes: the buoyancy driven regime and the convection-driven regime.
The transition between the two regimes is governed by the balance between the buoyancy flux of the plume,  $g'Q$, where $g'$ is effective gravity and $Q$ is volumetric flow rate, and the momentum flux of the entrained atmosphere.
For higher wind speeds, above around \SIrange{4}{5}{\meter\per\second} in \cref{fig:T_30_vs_wind,fig:T_10_vs_wind}, the horizontal momentum of the entrained air quickly dominates over the vertical momentum flux driven by buoyancy in the plume.
The transition between the two regimes is pushed to higher wind speeds as buoyancy is increased.
This can be seen when comparing the results from the dry simulations to the humid simulation, and it is also true when comparing \SI{30}{\celsius} and \SI{10}{\celsius}, as buoyancy is stronger in the colder atmosphere.
In the colder atmosphere, the safety distance is also reduced simply because the denser ambient air has more molecules per volume for the ammonia to dilute in.

In the buoyancy-driven regime, at low wind velocities, the plume is bent more with increasing wind speed. Increased bending means the plume reaches further before either being diluted or being lifted above the threshold height of \SI{3}{\meter}.
On the other hand, in the convection-driven regime, the buoyancy of the plume becomes negligible, and the ammonia is mainly transported passively in the horizontal direction by the wind.
While increased wind speed means a faster transport, it also leads to a faster dilution, and the result is that the plume travels a shorter distance.

If one neglects the influence of the ammonia on the velocity profile and assumes a constant wind speed, one gets a simple Gaussian plume with an ammonia concentration that is $n_{\nht} \propto Q/(ux^2)$ along the centerline, where $u$ is the wind speed.
In this case, solving for $x$ yields a safety distance that is proportional to $1/\sqrt{n_{\nht} u}$.
Although the velocity profile close to the ground is not constant, we nevertheless find that the safety distances at wind speeds above \SI{7}{\meter\per\second} are well approximated by $1/\sqrt{u}$, as indicated by the black dotted lines in \cref{fig:T_30_vs_wind,fig:T_10_vs_wind}.
On the other hand, the safety distance grows faster than $1/\sqrt{n_{\nht}}$ in the convection-driven regime.
This is likely because the lower concentrations are able to creep close to the ground where the velocity, and therefore dispersion, is smaller.

Comparing \cref{fig:T_30_vs_wind,fig:T_10_vs_wind} we see that the effect of fog formation is similar at \SI{30}{\celsius} and \SI{10}{\celsius}.
While the effect is stronger in the warmer atmosphere, the difference is less than one might expect based on the fact that the water content in the air is about \SI{350}{\percent} more in \SI{30}{\celsius} air compared to \SI{10}{\celsius} air, at the same relative humidity.
For example, at $u_\text{wind} = \SI{3}{\meter\per\second}$ and $T_\text{amb} = \SI{30}{\celsius}$, the increase in safety distance due to the absence of fog is \SI{41}{\percent}, \SI{54}{\percent},  and \SI{33}{\percent} for, respectively, 20000~ppm, 10000~ppm, and 5000~ppm, while for $T_\text{amb} = \SI{10}{\celsius}$, the increase for the same safety distances are \SI{16}{\percent}, \SI{33}{\percent},  and \SI{33}{\percent}.
Hence, considering ammonia-water fog formation can be important also in relatively colder weather.

\begin{figure}[htpb]
  \centering
  \includegraphics[width=0.48\textwidth]{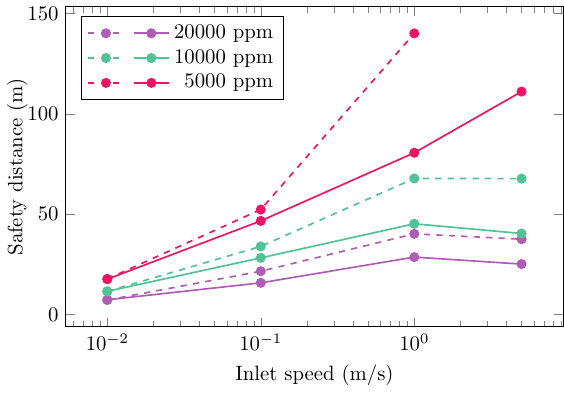}
  \caption{Safety distances as a function of ammonia inlet speed, $u_\text{in}$ for the reference case (\cref{tab:ref_case}) with $L = \SI{1}{\meter}$.  Solid lines have \SI{99.5}{\percent} relative humidity while dashed lines have \SI{0}{\percent} relative humidity. The black dotted line indicates the total length of the mesh and therefore the maximum length in the simulations.}
  \label{fig:uI}
\end{figure}

\subsubsection{Inlet velocity}
\Cref{fig:uI} shows the safety distances at 20000 ppm, 10000 ppm, and 5000 ppm with and without humidity as a function of ammonia inlet speed.
The inlet length is set constant at \SI{1}{\meter}.
The lower velocities on the scale of centimeters per second correspond to what one might expect from ammonia spills on dry ground, while the faster speeds are more akin to what can be expected from spills on water.
At $u_\text{in} = \SI{e-2}{\meter\per\second}$ the volumetric flux of ammonia is only \SI{0.01}{\meter\cubed\per\second}, and we are well within the convective regime.
In accordance with the previous, discussion, it is therefore no difference between the humid atmosphere and the dry atmosphere in this case.

As expected, the safety distances in \cref{fig:uI} increase as $u_\text{in}$ increases.
From the simple Gaussian model, one would expect the safety distances to increase by $\sqrt{10} \approx \SI{300}{\percent}$ when going from $u_\text{in} = \SI{e-2}{\meter\per\second}$ to $u_\text{in} = \SI{e-1}{\meter\per\second}$.
While this is almost true in the dry case, where the safety distances increase from \SI{7.2}{\meter}, \SI{11.6}{\meter}, and \SI{17.7}{\meter} to \SI{21.6}{\meter}, \SI{34.0}{\meter}, and \SI{52.3}{\meter}, it is less true in the humid case where the safety distances increase by only \SIrange{114}{164}{\percent}.
This difference can be attributed to the enhanced buoyancy flux associated with fog formation that lifts the plume to altitudes where the wind speed is higher, making the constant-wind profile assumption used in the convection analysis invalid.

As $u_\text{in}$ increases to \SI{1}{\meter\per\second}, the Gaussian approximation becomes poor also for the dry case, especially for the concentrations above 10000 ppm.
The reason for this is in part the increased vertical momentum associated with the increased inlet speed and in part the increased buoyancy flux associated with the increased volumetric flow.
While it is difficult to compare these two contributions in the dry case, it is clear by comparing the dry case and the humid case that buoyancy is of major importance in the latter case.
The vertical momentum flux at $u_\text{in} = \SI{1}{\meter\per\second}$ is the same, but the safety distances in the dry case are between \SI{40}{\percent} and \SI{74}{\percent} larger.

As the inlet speed is increased beyond \SI{1}{\meter\per\second}, the safety distances start to decrease because the vertical momentum launches the jet above the \SI{3}{\meter} mark.
In the limit $u_\text{in} \to \infty$, the safety distances will all go to zero for this reason.
From a safety perspective, the most serious scenarios are those with inlet speeds that give large mass flows without having too much vertical momentum.
Buoyancy becomes especially important under such circumstances, and by extension the buoyancy effect of \nht-\wat fog formation can significantly affect the results.

\begin{figure}[htpb]
  \centering
  \includegraphics[width=0.48\textwidth]{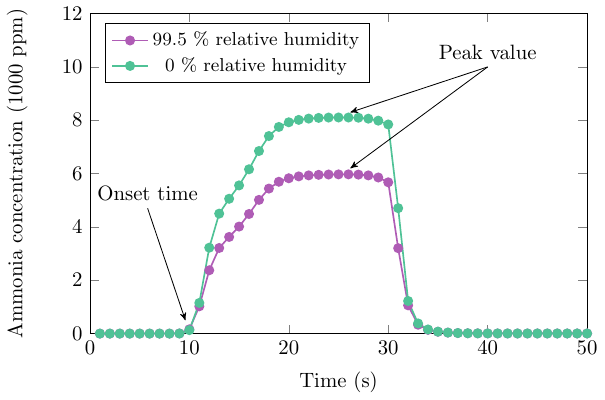}
  \caption{Ammonia concentration at $x= \SI{45}{\meter}$, $y= 0$ and $z = \SI{2}{\meter}$ as a function of time after the initialization of a \SI{3.39}{\kilogram\per\second} release lasting \SI{18}{\second}. The inlet side length is $L=\SI{0.5}{\meter}$, the wind speed is $u_\text{wind} = \SI{4.8}{\meter\per\second}$, the ambient temperature is $T_\text{amb} = \SI{8.5}{\celsius}$. The figure also indicates the definition of the onset time, which is where the concentration exceeds 50~ppm, and the peak values.}
  \label{fig:RS}
\end{figure}

\begin{figure}[htpb]
  \centering
  \includegraphics[width=0.48\textwidth]{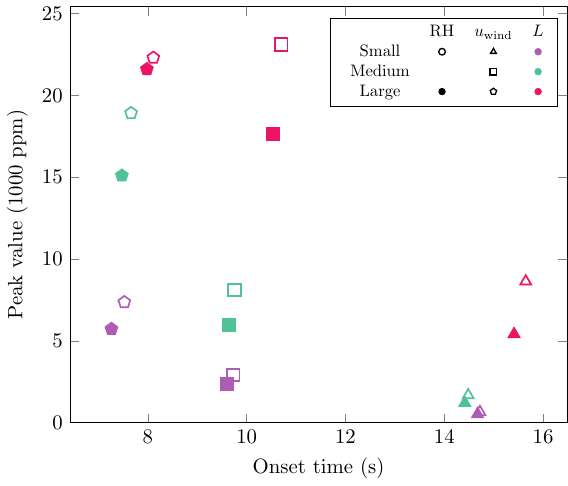}
  \caption{The onset time and peak value for transient puffs with different combinations of relative humidity ($\text{RH} \in \{\SI{0}{\percent}, \SI{99.5}{\percent}\}$), indicated by whether the mark is filled, wind speed ($u_\text{wind} \in \{\SI{3.1}{\meter\per\second}, \SI{4.8}{\meter\per\second}, \SI{6.5}{\meter\per\second}\}$), indicated by shape, and inlet side length, $L\in \{\SI{0.3}{\meter}, \SI{0.5}{\meter}, \SI{0.8}{\meter}\}$, indicated by color. For example, the filled green square shows the values corresponding to low relative humidity, meaning 0 \% and medium wind speed (\SI{4.7}{\meter\per\second}) and medium inlet side length (\SI{0.5}{\meter}). The ambient temperature is $T_\text{amb} = \SI{30}{\celsius}$ and the inlet mass flow is kept constant at \SI{3.39}{\kilogram\per\second}.}
  \label{fig:RS_puff}
\end{figure}

\begin{figure}[htpb]
  \centering
  \includegraphics[width=0.48\textwidth]{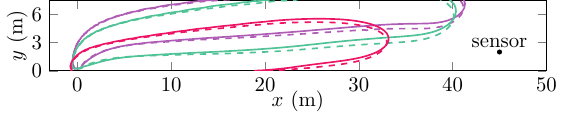}
  \caption{Contour lines of puffs at \SI{10}{\second} for the different scenarios with wind speed \SI{4.8}{\meter\per\second}. The colors indicate the same inlet size as in \cref{fig:RS_puff}, solid lines are with \SI{99.5}{\percent} relative humidity, and dashed lines are with \SI{0}{\percent} relative humidity.}
  \label{fig:RS_contour}
\end{figure}

\subsection{Red Squirrel Test 1F}
Finally, we consider a transient puff where a certain amount of liquid ammonia is spilled on water.
For concreteness we use the parameters from experiment RS-1F in Ref~\cite{dharmavaram2023}, where \SI{136}{\kilogram} ammonia was released for \SI{18}{\second} onto a \SI{6}{\centi\meter} deep bed of water, with an ambient air temperature of \SI{8.5}{\celsius}.
Of the \SI{136}{\kilogram}, \SI{61}{\kilogram} is evaporated upon contact with water, giving a vapor mass rate of \SI{3.39}{\kilogram\per\second}.
During the release, we set the inlet size constant and compute the inlet velocity according to $u_\text{in} = Q/L^2 = \SI{3.82}{\meter\cubed\per\second}/L^2$.
To model the bund we raise the inlet patch \SI{0.3}{\meter}.
A sensor that monitors ammonia concentration is placed \SI{45}{\meter} away from the source at a height of \SI{2}{\meter}.

Note that qualitative comparison with experiment has limited value in this case, as there are many important aspects we do not include in the present model.
For example, the evaporation rate depends on how ammonia mixes with water. 
There will be fast evaporation over a small region in the beginning, but this will change as the water cool and become more mixed with ammonia.
This will depend on how quickly ammonia is able to spread in the shallow bed of water.
Moreover, Reynolds-averaged transport equations are not well suited to study transient puffs.
To understand how the ammonia concentration measurement at a single point changes in time, one should instead perform large eddy simulations, include the interaction between the refrigerated ammonia and water and preferably have an accurate description of the dynamic ambient wind.
The aim here is instead to study how the measurements at such experiments are influenced by different parameters.

\Cref{fig:RS} shows the concentration at the ammonia sensor as a function of time for two different relative humidities.
The figure also shows the \emph{onset time}, which we define to be the time when the sensor first reads a value above 50~ppm, and the \emph{peak value}, which is the maximum concentration during the puff.
\Cref{fig:RS_puff} shows the peak value and onset time for various combinations of relative humidity, wind speed, and ammonia inlet size.
The contour lines at 10\,000~ppm for the cases with $u_\text{wind}=\SI{4.8}{\meter\per\second}$ at \SI{10}{\second} is shown in \cref{fig:RS_contour}.
Even though the temperature is only \SI{8.5}{\celsius}, and the Froude number is larger than for the previous cases, the \nht-\wat fog still has a noticeable impact on the results.

The onset time is almost entirely determined by the wind speed and is around $\SI{45}{\meter}/u_\text{wind}$.
The variations in onset time due to $L$ and RH can be attributed to the fact that wind speed is lower near the ground.
That is, a lower value of $L$ or a higher value of RH will lead to a plume that is lifted higher and therefore feels a stronger ambient wind speed.
However, this does not always mean a shorter onset time, because a smaller $L$ also means more vertical momentum that must be converted to horizontal momentum.
Thus, a smaller $L$ generally means a slower horizontal travel in the beginning and a faster horizontal travel after the horizontal momentum from the entrained air starts to dominate.
This explains the non-monotonic dependence of onset time on $L$ in the lower wind speed \cref{fig:RS_puff}.

On the other hand, the dependence of peak value on plume height above the sensor is monotonic as expected.
All three factors, wind speed, inlet size, and humidity, affect the peak values.
Although the inlet speed is up to \SI{42}{\meter\per\second}, and despite a low ambient temperature of only \SI{8.5}{\celsius}, the effect of fog-induced buoyancy is still significant.
This is because even a small relative lift in plume height, as shown in \cref{fig:RS_contour}, can strongly affect the ammonia concentration at the sensor.

By looking at the wind speed dependence in \cref{fig:RS_puff}, it is clear that we are mostly in the buoyancy-driven regime. 
The effect of fog-induced buoyancy in this regime mostly increases or stays constant.
The exception may be the combination of the highest wind speed and the lowest inlet speed, marked by the red pentagons in \cref{fig:RS_puff}.
Here the effect of humidity is relatively small, which indicates that the system has entered the convection-driven regime. 
This is consistent with the fact that the peak value has been reduced in the dry case compared to the lower wind speed marked by the red squares.

\section{Conclusions}\label{sec:conclusion}
We have implemented a CFD model for ammonia dispersion that successfully integrates ammonia-water fog formation and its effect on buoyancy.
The strongly exothermic reaction that takes place as ammonia mixes with water is included using a concentration-dependent heat of formation in the numerical model, while the phase change rate is chosen to facilitate rapid equilibrium through a composition equilibrium correlation function for \nht-\wat mixtures.
We validate the implementation against the full multiparameter EoS and find excellent agreement.

By applying the CFD model to ammonia dispersion under various conditions we find a significant impact of buoyancy from the \nht-\wat fog formation, even in modest temperatures below \SI{10}{\celsius}.
It is therefore important to include this effect when developing models to predict ammonia dispersion.
In the absence of the full CFD framework that is often too computationally demanding. Simpler plume dispersion models must employ empirical constants to describe, for example, the entrainment of ambient air.
If these parameters do not take into account how ambient humidity will affect the buoyancy, they can potentially be fitted to experiments performed on a humid day and fail to predict how far lethal amounts of \nht can spread on a dry day by more than \SI{50}{\percent}.

The spreading of ammonia can be separated into a buoyancy-driven regime, where buoyancy and/or vertical momentum is able to lift the plume, and a convection-driven regime, where the plume mostly follows the ground due to a strong transverse wind.
Ammonia spreads the furthest in the transition region between these two regimes, since increasing the wind further reduces the distance due to enhanced entrainment.
We find that fog-induced buoyancy has a significant impact on the result in this region, and the worst-case case spreading distance is reduced by as much as \SI{25}{\percent} on a humid day compared to a dry day without fog formation.
We also implemented a pure \wat fog model to study whether this first approximation can be adequate.
The result is qualitatively similar, but the quantiative effect is significantly smaller.

The buoyancy effect of \nht-\wat fog formation is not a linear function of \nht concentration.
One might therefore expect the results to be modified when going to a more sophisticated turbulence model that can capture the large eddies of dispersing plumes.
These eddies are smeared out in the Reynolds-averaging procedure that is used in the current work.
Our implementation of the \nht-\wat fog, including the mixing enthalpy, can readily be used also in for example large eddy simulations.
For future work, it would be interesting to study the effect of large eddies.

While the results presented here unequivocally show that \nht-\wat fog formation is important for understanding the physics of ammonia dispersion, it is also clear that the magnitude of the effect is sensitive to other parameters, like the mass and momentum flux of the released ammonia.
It would therefore be interesting to apply the model to realistic loss of containment scenarios where both the mass and momentum flux vary in time and space.
While the model presented here can predict the dispersion of the fog and gaseous plume, much still remains when it comes to determining the relevant input parameters to realistic LOC scenarios.
This requires a coupled hydrodynamic and thermodynamic model of liquid \nht spreading on the ground or in water.

\section*{Acknowledgments}
This work was funded by the Research Council of Norway through the MaritimeNH3 project, project number 328679.

\bibliographystyle{elsarticle-num}
\bibliography{literature.bib}

\end{document}